\documentstyle[preprint,aps]{revtex}

\begin{document}
\title{Anomalous Band Structure in Odd-Odd Nuclei\\ with the
Quadrupole-Quadrupole Interaction}
\author{M.S. Fayache$^1$ and L. Zamick$^2$\\
(1) D\'{e}partement de Physique, Facult\'{e} des Sciences de Tunis\\
Tunis 1060, Tunisia\\
\noindent (2) Department of Physics and Astronomy, Rutgers University\\
        Piscataway, New Jersey 08855\\}
\date{\today}
\maketitle

\begin{abstract}

We perform shell model calculations in odd-odd nuclei using a
quadrupole-quadrupole interaction with single-particle splittings
chosen so as to obtain the $SU(3)$ results. 
Elliott had shown that such an interaction gives rotational bands for
which the energies go as $I(I+1)$. This certainly is true for
even-even and for odd-even or even-odd nuclei with $K\neq 1/2$. We
have looked at odd-odd nuclei e.g. $^{22}Na$ and found somewhat
different behaviour. In $^{22}Na$ the $I=1^+_1~T=0$ and $I=0^+_1~T=1$
states are degenerate, and a rotational band built on the
$I=0^+_1~T=1$ state behaves in a normal fashion. For the $I=1^+_1~T=0$
band however, we find that the energy is given by
$E(I)-E(1^+_1)=AI(I+1)$. This differs from the `normal' behaviour
which would be $E(I)-E(1^+_1)=AI(I+1)-2A$. 

\end{abstract}

\section{Introduction}
In the rotational model the formula for the energy of a state in a
rotational band with total angular momentum $I$ is given by \cite{bm}

\begin{equation}
E_I=E_0+\frac{\hbar^2}{2\cal J}\left[ I(I+1)+\delta_{K,1/2}
a(-1)^{I+1/2}(I+1/2)\right]
\end{equation}

\noindent where $a$ is the decoupling parameter given by
$a=-\langle K=1/2~|~ J_{+}~|~ \overline{K=1/2} \rangle $ and where if
$|K \rangle=\sum_{j}C_{j,k}\phi_{j,k}$ then
$|\bar{K}\rangle=\sum_{j}C_{j,k}(-1)^{j+k}\phi_{j,-k}$.

\noindent For even-even nuclei, and for odd-even and even-odd nuclei
with $K\neq 1/2$, one gets the familiar $I(I+1)$ spectrum \cite{Elliott}. 

It is generally thought that the Elliott $SU(3)$ model also gives an
$I(I+1)$ spectrum. This has been discussed most explicitly in the
context of even-even nuclei. The $SU(3)$ results also give the more
complex $K=1/2$ behaviour where the decoupling parameter $a$ has a
value corresponding to that obtained from an asymptotic Nilsson wave
function. This will be discussed briefly in section II. But the main
thrust of this work will be to show that for odd-odd nuclei one
obtains in certain cases deviations from the above formula. 

We have performed shell model calculations with all possible
configurations in a given major shell using the interaction
$\sum_{i<j}Q(i)\cdot Q(j)$ where, in order to get Elliott's $SU(3)$
results we must also add single-particle splittings, e.g. in the
$1s-0d$ shell we have $\epsilon_{0d}-\epsilon_{1s}=18\bar{\chi}$ and in the 
$1p-0f$ shell we have $\epsilon_{0f}-\epsilon_{1p}=30\bar{\chi}$, where 
$\bar{\chi}=\frac{5b^4\chi}{32\pi}$ with $b$ the oscillator length
parameter $(b^2=\frac \hbar {m\omega })$.

As has been previously noted \cite{fay1,moya}, we use the
$\vec{r}$-space $Q \cdot Q$ interaction rather than the mixed
$\vec{r}$ and $\vec{p}$-space one. With such an interaction 2/3 of the
above single-particle splitting comes from the $i=j$ part of $Q \cdot
Q$ and 1/3 from the interaction of the valence particle with the core. 

\section{A Brief Look at $K=1/2$ Bands}

Let us be specific and discuss $^{19}F$ and $^{43}Sc$. We consider in
each case three valence nucleons beyond a closed shell. In $^{19}F$
the particles are in the $1s-0d$ shell, whereas in $^{43}Sc$ they are
in the $1p-0f$ shell. The energy levels of the lowest bands are given
in Table I for the two cases. The results for the two nuclei are
striking but different. In $^{19}F$, the lowest state is a $I=1/2^+$
singlet, and at higher energies we get degenerate pairs
$(3/2^+,5/2^+)$, $(7/2^+,9/2^+)$, $(11/2^+,13/2^+)$. In $^{43}Sc$ the
ground state is degenerate, and the degenerate pairs are 
$(1/2^+,3/2^+)$, $(5/2^+,7/2^+)$,..., $(17/2^+,19/2^+)$.

If we look at the rotational formula, we find that these results are
consistent with a decoupling parameter $a=+1$ for $^{19}F$  and $a=-1$
for $^{43}Sc$. It is easy to show that these are precisely the results
one obtains with asymptotic Nilsson wave functions. In both cases the
odd particle will be in a $\Lambda=0~\Sigma=1/2$ state in the
asymptotic limit. From the definition of $\bar {K}$, the state
$|\bar{\Lambda=0~\Sigma=1/2}\rangle$ can be shown to be equal to
$-(-1)^{\pi}|\Lambda=0~\Sigma=-1/2 \rangle$ where $\pi$ is $(+)$ for
an even-parity major shell and $(-)$ for an odd-parity one. Hence:\\
\[a=(-1)^{\pi}\langle \Sigma=+1/2~|~ J_{+}~|~
\Sigma=-1/2\rangle=(-1)^{\pi}\] 

\noindent It has long ago been noted by Bohr and Mottelson \cite{bm}
that $a=+1$ corresponds to weak coupling of the odd particle to
$I=0^+,~2^+,~4^+,...$ states, whereas $a=-1$ corresponds to weak
coupling to $I=1,~3,~5,...$ states. 
It should be emphasized that the results in Table I are not the
realistic ones -they represent the asymptotic extremes. 

At any rate, we have shown that the $Q \cdot Q$ interaction gives the
same results for these two $K=1/2$ bands as does the rotational
formula with asymptotic Nilsson wave functions.

\section{Odd-Odd Nuclei e.g. $^{22}Na$}
\subsection{The Energy Spectra}

In table II we show a fairly detailed list of energy levels for the
odd-odd nucleus $^{22}Na$ obtained with the $Q\cdot Q$ interaction. We
show $T=0$ and $T=1$ states in separate columns. We have
underlined $T=0$ and $T=1$ rotational bands, and will now discuss them
in more detail. We use the same parameters as in $^{19}F$ just to
bring out some similarities. If one is interested in a best fit, one
should of course have an $A$ dependence in $\chi$.

Note that the ground state consists of two degenerate states, one with
$I=1^+~T=0$ and the other with $I=0^+~T=1$. Both states have $L=0$ and
the simple spin-independent interaction gives the same energy for
$S=0$ and $S=1$. Let us first look at the $T=1$ states. The ground
state is $I=0^+$. The $2^+$ state is at 1.588 and is doubly
degenerate. If we follow the rotational sequence $I=0^+,~2^+,~4^+,...$
we see a simple rotational behaviour:\\

\[E(I)-E(0^+_1)=AI(I+1)~~~~~~~\left [A=\frac{\hbar^2}{2\cal
J}=E(2^+)/6\right ]\] 

\noindent There is nothing new here.

We next look at the $T=0$ states. The lowest state has $I=1^+$ (it is
degenerate with the lowest $I=0^+$ $T=1$ state). If we follow the
underlined states we have a $2^+$ at 1.588 $MeV$, $3^+$ at 3.177
$MeV$, $4^+$ at 5.293 $MeV$, $5^+$ at 7.941 $MeV$ until we reach
$10^+$ at 29.117 $MeV$.

The energy levels of $I=2^+,~3^+,...,~9^+,~10^+$ are given by 

\[E^*(I)\equiv E(I)-E(1^+_1)=AI(I+1)~~~~~~~\left [A=\frac{\hbar^2}{2\cal
J}=E(2^+)/6\right ]\] 

\noindent At first sight there would appear to be nothing wrong. 
But remember that $E^*(I)$ is the energy of
a state of angular momentum $I$ for which the $I=1^+$ state has been
set to zero energy. If we put $I=1$ into the above formula we would
get $E^*(1)=2A$.

To put it in a better way, the rotational formula at the beginning of
this paper (Eq. (1)) would yield

\[E(I)-E(1^+)=AI(I+1)-2A\]

\noindent However, the results that we obtain are 

\begin{eqnarray*}
E^*(I)=E(I)-E(1^+) & = & AI(I+1)~~~~~~I\neq 1\\
& = & 0~~~~~~~~~~~~~~~~~I=1\\
\end{eqnarray*}

\noindent Thus, for the case of $T=0$ states in odd-odd nuclei we get
a difference between the rotational formula and the $SU(3)$ limit.

\subsection{The $B(E2)$ Values for $T=0~\rightarrow T=0$ Transitions} 

To clarify the structure of these bands, we performed calculations of
$B(E2)$ values for various $T=0~\rightarrow T=0$ transitions up to
$I=4$. They are shown in Tables III and IV, where we introduced a
small spin-orbit splitting in order to remove the degeneracies as our
shell model code does not handle transitions involving degenerate
states very well. Note that with bare $E2$ charges $e_p=1,~e_n=0$ we
obtain $B(E2:1_1^+~T=0 \rightarrow 2_1^+~T=0)=34.9~e^2fm^4$. This is
quite large, and in our opinion justifies treating the $I=1^+$ state
as a member of the band. Actually, if we used the usual effective
charges $e_p=1.5,~e_n=0.5$, the $B(E2)$ value would increase four-fold 
(i.e. to about 140 $e^2fm^4$). Note also that the cross-over
transition $I=1^+_1~T=0 \rightarrow I=3^+_2~T=0$ at 3.2 $MeV$ is
zero. This is consistent with the $I^+_1$ state being $L=0~S=1$ and the
$I=3^+_2$ state being $L=3~S=1$. One cannot connect from $L=0$ to
$L=3$ via the $E2$ operator. There is some strength to a lower $3^+$
state which is {\em not} a member of the rotational band
($B(E2)=6.15~e^2fm^4$). That $3^+$ state {\em must} be $L=2~S=1$.

Our work suggests that the rotational model formula requires an
additional term for odd-odd nuclei in order to be consistent with the
$SU(3)$ results \cite{Elliott}. 
We gain further insight by examining the degeneracies associated with
the $T=0$ underlined states of Table II, i.e. those with energy
$AI(I+1)$. The even $I$ states up to $I=8$ are doubly degenerate
whereas the others are singlets. This suggests that there are two
bands for which the states with the same $I$ values are
degenerate. One band is a $K=2$ band with all values of $I$ from 2 to
10, and there is nothing anomalous about it. The other band consists
of states of angular momentum 1,2,4,6 and 8. For the latter band, the 
{\em orbital} angular momentum of the states are 0,2,4,6 and 8
respectively, and they all have $S=1$. Their energies can be fit to
the formula $E^*(I)=BL(L+1)$ rather than $AI(I+1)$.

\acknowledgements

We thank Ben Bayman for clarifying remarks about the rotational model
for odd-odd nuclei. This work was supported by a Department of Energy
Grant No. DE-FG02-95ER40940. M.S. Fayache would like to thank the
Nuclear Theory Group in the Department of Physics at Rutgers
University for its generous hospitality, and kindly acknowledges
travel support from the Facult\'{e} des Sciences de Tunis,
Universit\'{e} de Tunis, Tunisia.

\nopagebreak

\nopagebreak
\begin{table}
\caption{Energy Levels (in $MeV$) of Excited States Corresponding to
the $K=1/2$ Ground State Bands in $^{19}F$ and $^{43}Sc$ 
with the $-\chi Q\cdot Q$ Interaction.}
\begin{tabular}{ccccc}
\multicolumn{2}{c}{$^{19}F$\tablenotemark[1]} & &
\multicolumn{2}{c}{$^{43}Sc$\tablenotemark[2]}\\
\tableline
$I^{\pi}$ & $E^*$ & & $I^{\pi}$ & $E^*$\\
 $(\frac{1}{2})^+$ & 0 & & $(\frac{1}{2})^-$ & 0\\
 $(\frac{3}{2})^+$ & 1.588 & & $(\frac{3}{2})^-$ & 0\\
 $(\frac{5}{2})^+$ & 1.588 & & $(\frac{5}{2})^-$ & 0.679\\
 $(\frac{7}{2})^+$ & 5.295 & & $(\frac{7}{2})^-$ & 0.679\\
 $(\frac{9}{2})^+$ & 5.295 & & $(\frac{9}{2})^-$ & 1.900\\
 $(\frac{11}{2})^+$ & 11.118 & & $(\frac{11}{2})^-$ & 1.900\\
 $(\frac{13}{2})^+$ & 11.118 & & $(\frac{13}{2})^-$ & 3.664\\
                    &        & & $(\frac{15}{2})^-$ & 3.664\\
                    &        & & $(\frac{17}{2})^-$ & 5.971\\
                    &        & & $(\frac{19}{2})^-$ & 5.971\\
\end{tabular}
\tablenotetext[1]{For $^{19}F$ we use $\chi=0.1841$
($\bar{\chi}=0.0882$)}
\tablenotetext[2]{For $^{43}Sc$ we use $\chi=0.0294$ ($\bar{\chi}=0.0218)$}
\end{table}

\begin{table}
\caption{The Energy Levels (in $MeV$) of $^{22}Na$ Calculated 
with the $-\chi Q\cdot Q$ Interaction$^a$}
\begin{tabular}{ccc}
$I^{\pi}$ & $T=0$ States &  $T=1$ States\\
\tableline
    $0^+$ &    8.999	&   \underline{0.000}\\
          &   12.176	&   2.647\\		
          &   12.176	&   8.999\\		
          &   13.235    &   9.000\\	  
          &   16.410	&  12.176\\
          &             &\\
    $1^+$ & \underline{0.000} & 2.647\\
          &    1.588    &       8.999\\
          &    1.588    &       8.999\\
          &    2.647    &      10.059\\
          &    9.000    &      10.059\\
          &             &\\
    $2^+$ & \underline{1.588} & \underline{1.588}\\
          &    1.588    &       1.588\\
          &    3.176    &       2.647\\
          &    8.999 	&       5.294\\
          &   10.059	&       9.000\\
          &             &\\
    $3^+$ &    1.588	&       3.176\\
          &    1.588	&       5.293\\
          & \underline{3.177} &10.058\\
	  &    5.294	&      10.058\\
	  &    5.294	&      11.646\\
          &             &\\
    $4^+$ &    3.176    &      \underline{5.293}\\
          & \underline{5.293} & 5.293\\
          &    5.293    &       5.293\\
          &    7.941    &      10.059\\
          &   11.647    &      11.647\\
          &             &\\
    $5^+$ &   5.294	&       7.941\\
          &   5.294	&      10.059\\
          & \underline{7.941} & 13.763\\
          &  10.059 	&       13.763\\
          &  11.118	&       13.763\\
          &             &\\
    $6^+$ &  7.941	&       10.059\\
          & \underline{11.117} &  \underline{11.117}\\
          & 11.117	&       11.118\\
          & 14.824	&       16.412\\
          & 16.411	&       16.412\\
          &             &\\
    $7^+$ & 11.117	& \underline{14.823}\\
          & 11.117	&       16.940\\
          & \underline{14.823} & 19.587\\
          & 16.941	&        19.587\\
          & 19.058      &        19.587\\
          &             &\\
    $8^+$ & 14.823	&      16.941\\
          & \underline{19.058} & \underline{19.058}\\
          & 19.059	&      19.059\\
          & 22.763	&      22.767\\
          & 23.292	&      23.293\\
          &             &\\
    $9^+$ & 19.058	&  \underline{23.822}\\
          & 19.058	&      25.939\\
          & \underline{23.822} &  26.470\\
          & 25.940      &      27.527\\
          & 26.469	&      27.527\\
          &             &\\
   $10^+$ & 23.823	&      25.942\\
          & \underline{29.117}  & \underline{29.117}\\
          & 30.705	&      30.706\\
          & 32.293	&      32.294\\
          & 33.881	&      32.294\\
\end{tabular}
\tablenotetext[1]{In this table and in the following tables, the same
value of $\chi$ (and of $\bar{\chi}$) was used for $^{22}Na$ as for
$^{19}F$.} 
\end{table}

\begin{table}
\caption{Calculated $B(E2)$ from the Ground State in $^{22}Na$ with 
the $-\chi Q \cdot Q$ Interaction.}
\begin{tabular}{cc}
\multicolumn{2}{c}{$I=1_1^+~T=0~\rightarrow I=2^+~T=0$}\\
$E^*(I=2^+,T=0)$ & $B(E2)~(e^2 fm^4)$\\
1.591 & 34.89\\
1.598 & 4.31\\
3.199 & 0.00\\
8.995 & 0.00\\
10.054 & 0.00\\
\tableline
\multicolumn{2}{c}{$I=1_1^+~T=0~\rightarrow I=3^+~T=0$}\\
$E^*(I=3^+,T=0)$ & $B(E2)~(e^2 fm^4)$\\
1.570 & 6.15\\
1.586 & 48.74\\
3.183 & 0.00\\
5.306 & 0.00\\
5.320 & 0.00\\
\end{tabular}
\end{table}

\begin{table}
\caption{Calculated $B(E2)$ Between Excited States in $^{22}Na$ with 
the $-\chi Q \cdot Q$ Interaction.}
\begin{tabular}{ccc}
\multicolumn{3}{c}{$I=2^+~T=0~\rightarrow I=3^+~T=0$}\\
$E^*(I=2^+,T=0)$ & $E^*(I=3^+,T=0)$ & $B(E2)~(e^2 fm^4)$\\
1.591 & 1.570 & 0.00\\
1.591 & 1.586 & 13.73\\
1.591 & 3.183 & 3.18\\
      &       &\\
1.598 & 1.570 & 13.5\\
1.598 & 1.586 & 0.00\\
1.598 & 3.183 & 26.57\\
      &       &\\
3.199 & 1.570 & 1.44\\
3.199 & 1.586 & 0.20\\
3.199 & 3.183 & 0.00\\
      &       &\\
\tableline
\multicolumn{3}{c}{$I=2^+~T=0~\rightarrow I=4^+~T=0$}\\
$E^*(I=2^+,T=0)$ & $E^*(I=4^+,T=0)$ & $B(E2)~(e^2 fm^4)$\\
1.591 & 3.161 & 1.41\\
1.591 & 5.296 & 30.60\\
1.591 & 5.299 & 12.14\\
      &       &\\
1.598 & 3.161 & 11.03\\
1.598 & 5.296 & 1.14\\
1.598 & 5.299 & 25.00\\
      &       &\\
3.199 & 3.161 & 0.00\\
3.199 & 5.296 & 0.59\\
3.199 & 5.299 & 5.83\\
      &       &\\
\tableline
\multicolumn{3}{c}{$I=3^+~T=0~\rightarrow I=4^+~T=0$}\\
$E^*(I=3^+,T=0)$ & $E^*(I=4^+,T=0)$ & $B(E2)~(e^2 fm^4)$\\
1.570 & 3.161 & 40.10\\
1.570 & 5.296 & 0.15\\
1.570 & 5.299 & 2.68\\
      &       &\\
1.586 & 3.161 & 5.03\\
1.586 & 5.296 & 4.37\\
1.586 & 5.299 & 1.73\\
      &       &\\
3.183 & 3.161 & 0.00\\
3.183 & 5.296 & 2.48\\
3.183 & 5.299 & 24.61\\
      &       &\\
\tableline
\multicolumn{3}{c}{$I=3^+~T=0~\rightarrow I=5^+~T=0$}\\
$E^*(I=3^+,T=0)$ & $E^*(I=5^+,T=0)$ & $B(E2)~(e^2 fm^4)$\\
1.570 & 5.278 & 20.31\\
1.570 & 5.289 & 0.36\\
1.570 & 7.945 & 0.00\\
      &       &\\
1.586 & 5.278 & 8.74\\
1.586 & 5.289 & 36.00\\
1.586 & 7.945 & 0.00\\
      &       &\\
3.183 & 5.278 & 4.53\\
3.183 & 5.289 & 0.20\\
3.183 & 7.945 & 25.09\\
\end{tabular}
\end{table}

\end{document}